\documentstyle[preprint,aps]{revtex}

\newcommand{\mathR}{{\rm I\! R}}                

\newtheorem{thm}{Theorem}
\newtheorem{lem}{Lemma}

\begin{document}
\draft
\title{\hfill {\rm IMPERIAL/TP/95--96/1}\\[0.8cm]
	The Symplectic Geometry of a\\ Parametrized Scalar Field on a
Curved Background}

\author{P.~H\'{a}j\'{\i}\v{c}ek\thanks{email: hajicek@butp.unibe.ch}}
\address{Institute for Theoretical Physics\\
		University of Bern\\
		Sidlerstrasse 5, CH-3012 Bern, Switzerland}

\author{C.J.~Isham\thanks{email: c.isham@ic.ac.uk}}
\address{Theoretical Physics Group,
		Blackett Laboratory\\
		Imperial College of Science, Technology \& Medicine\\
		South Kensington, London  SW7 2BZ, U.K.}

\date{October, 1995}
\maketitle

\begin{abstract}
We study the real, massive Klein-Gordon field on a $C^\infty$
globally-hyperbolic background space-time with compact Cauchy
hypersurfaces.  In particular, the parametrization of this system as
initiated by Dirac and Kucha\v{r} is put on a rigorous basis.
The discussion is focussed on the structure of the set of spacelike
embeddings of the Cauchy manifold into the space-time, and on the
associated $e$-tensor density bundles and their tangent and cotangent
bundles.  The dynamics of the field is expressed as a set of
automorphisms of the space of initial data in which each pair of
embeddings defines one such automorphism. Using these results, the
extended phase space of the system is shown to be a weak-symplectic
manifold, and the Kucha\v{r} constraint is shown to define a smooth
constraint submanifold which is foliated smoothly by the constraint
orbits. The pull-back of the symplectic form to the constraint
surface is a presymplectic form which is singular on the tangent
spaces to the constraint orbits.  Thus, the geometric structure of
this infinite-dimensional system is analogous to that of a
finite-dimensional, first-class parametrized system, and hence many
of the results for the latter can be transferred to the
infinite-dimensional case without difficulty.
\end{abstract}
\pacs{}

\section{Introduction}
\label{Sec:intro}
The long history of studies of quantum field theory in a background
space-time peaked sharply in the seventies (for reviews from that
era, see \cite{dewitt,isham-texas}) following Hawking's discovery of
the quantum radiation produced by a black hole \cite{Hawking}.  In
those days, the main aim was to find a direct quantization of the
true physical degrees of freedom of the system. However, much
earlier, Dirac had reformulated this system in a parametrized form
so that it could be used as a model for general relativity proper
\cite{K45}.  The idea is to treat embeddings of Cauchy hypersurfaces
in the space-time as additional degrees of freedom of the system.
Together with their conjugate momenta and the Cauchy data of the
scalar field on the embeddings, they define an extended phase space.
To retrieve the original dynamics one has to impose constraints.
This procedure was studied in some detail in the references
\cite{hyperspace,K47} and \cite{I+K}; a shorter exposition can be
found in \cite{K33}. The resulting system can be classified as a
`first-class parametrized system'.

	A method of quantizing first-class parametrized systems was
initiated by Dirac; in particular, he studied a system of
massive particles in Minkowski space-time \cite{dirac}. A
generalization of this method to any {\em finite}-dimensional
first-class parametrized system was given by H\'{a}j\'{\i}\v{c}ek
\cite{timelevels} using a combination of Dirac's ideas  with the group
quantization method of Isham \cite{isham} and the algebraic
quantization method of Ashtekar \cite{blue-book}. In what follows,
this generalization will be referred to as the `perennial formalism'.

	If one is interested in extending the perennial formalism to
infinite-dimensional systems, the scalar field on a fixed space-time
offers itself naturally as a well-understood and much-discussed
model. However, the perennial formalism is based on a geometrical
form of Hamiltonian dynamics, whereas the existing formulations
of the parametrized scalar field are non-geometrical.  The main
purpose of the present paper is to recast the classical theory of a
parametrized scalar field in a fixed background into the geometrical,
Hamiltonian form for infinite-dimensional systems developed by
Marsden and collaborators (see \cite{C+M}).  The ensuing results are
of interest in themselves and are also used in the accompanying
paper \cite{H+I} dealing with the application of the perennial
formalism to the quantum theory of a field propagating on a curved
background.

	Marsden's techniques have been applied to a variety of systems,
including general relativity itself\cite{F+M}. In all these
examples, the extended phase space is an (open) subset of a linear
space, and hence the tangent space at any point in the phase space
can be naturally identified with this linear space. The resulting
significant simplifications have been thoroughly exploited in the
literature (for example, see \cite{F+M}). However, in our case, the
space of embeddings is a genuine manifold, not just a subspace of a
linear space. Of course, the rough idea of how the theory is to be
applied in general to such cases is well-known \cite{C+M}, but---as
we shall see---the specific system of interest to us possesses some
crucial additional structure that is very helpful in the detailed
analysis.

	The plan of the paper is as follows. In section
\ref{Sec:dynamics}, the theory of the Cauchy problem of
a hyperbolic, partial differential equation (as reviewed, for
example, in \cite{F+M}) is applied to the dynamics of a massive
scalar field in a globally hyperbolic space-time. The Cauchy
hypersurfaces are assumed to be compact, but it seems likely
that---if desired---the proofs (which are relegated to the Appendix)
could be adapted to deal with asymptotically flat Cauchy
hypersurfaces.  The set of all Cauchy data is given the structure of
a topological vector space $\Gamma_{\phi}$, and an automorphism of
this space is associated with each oriented pair of embeddings of
the Cauchy manifold in the space-time, thereby producing a rather
generalised concept of `time evolution'.

	In section \ref{Sec:extended_phase_space}, we extend the phase
space from $\Gamma_{\phi}$ to $\Gamma_{\phi}\times T^*{\cal E}$,
where $\cal E$ is the space of embeddings introduced by Kucha\v{r}
in \cite{hyperspace}. This way of parametrizing the system was
discussed in detail by Isham and Kucha\v{r} \cite{I+K}. The
structure of the $e$-tensor density bundles over $\cal E$ is
described, and their tangent and cotangent bundles are studied; this
is where we lay the foundation for the subsequent mathematical
developments.  Our approach to the resulting infinite-dimensional
Hamiltonian system is based on the geometrical ideas of Chernov,
Fischer, and Marsden \cite{C+M,F+M}.  The constructions of the phase
space manifold, the symplectic structure, and the Poisson brackets,
are all described explicitly.

	In section \ref{Sec:constraint_manifold}, we show that (i) the
constraint set is a submanifold of the phase space; (ii) the
constraint orbits are submanifolds of the constraint surface; and
(iii) the vector space defined by the right hand side of the
evolution equation coincides with the tangent space of a constraint
orbit.  For the model we study, the proofs (which are adapted from
\cite{F+M}) are straightforward. The resulting structure is
analogous to a high degree to that of a finite first-class
parametrized system, and hence the application of the perennial
formalism is relatively unproblematic.

\section{Field dynamics}
\label{Sec:dynamics}
We are interested in the theory of a relativistic field propagating
on a fixed background space-time $\cal M$. The associated dynamical
equation defines an evolution map between Cauchy data along two
arbitrary Cauchy hypersurfaces. We shall need various properties of
these maps that can be derived from results concerning the Cauchy
problem for linear hyperbolic systems as described---for example---in
\cite{F+M,H+E}.

	The following properties of the space-time $({\cal M},g)$ will be
assumed:
\begin{enumerate}
 \item The space-time $\cal M$ is equipped with a $C^{\infty}$
differential structure.  In particular, this means that
diffeomorphisms do not mix the different Sobolev structures that
will be placed on various function spaces associated with $\Sigma$
and $\cal M$.

 \item The Lorentzian metric $g$ is such that the pair $({\cal M},g)$ is
globally hyperbolic. Thus the four-manifold $\cal M$ is necessarily
diffeomorphic to $\Sigma\times\mathR$ where the three-manifold
$\Sigma$ is a model for any Cauchy hypersurface in $\cal M$.

\item The three-manifold $\Sigma$ is compact. This assumption is
made for the sake of simplicity, but we expect that similar results
can be obtained for asymptotically flat space-times using
analogous---but more laborious---methods; see \cite{F+M}).
\end{enumerate}

	The relativistic wave equation of interest is the Klein-Gordon
equation for a real scalar field:
\begin{equation}
   	|\det g|^{-1/2} \partial_\mu(|\det g|^{1/2}g^{\mu\nu}
	\partial_\nu\phi) + m^2\phi = 0           		\label{K-G}
\end{equation}
where the real, non-negative constant $m$ is the mass
parameter.

	A central role in the theory is played by $C^{r+1}$ ($r>2$)
embeddings $X:\Sigma\rightarrow {\cal M}$ that are spacelike with
respect to $g$. We shall refer to any such $X$ simply as an
`embedding', and denote by $\text{Emb}_g(\Sigma,{\cal M})$ the
space of all such (see \cite{I+K} and \cite{hyperspace}). Each
embedding $X$ determines a positive-definite $C^r$-metric $\gamma$
on $\Sigma$ as the pull-back $X^*g$ of $g$ by $X$; {\em i.e.},
in local coordinates on both $\Sigma$ and $\cal M$,
\begin{equation}
	\gamma_{ab}(x):=g_{\mu\nu}(X(x))\, X^\mu,_{a}(x)\, X^\nu,_{b}(x).
							\label{Def:gamma}
\end{equation}
The map $X$ also determines a future-oriented, unit normal
vector $n(X(x))$ at each point of the hypersurface $X(\Sigma)$ in
$\cal M$.

	Any embedding $X\in\text{Emb}_g(\Sigma,{\cal M})$ defines a
`Cauchy datum' for $\phi$ along the hypersurface $X(\Sigma)$. This
is a pair $(\varphi,\pi)$ of fields on $\Sigma$, where the scalar
$\varphi$ and the density (of weight $w=1$) $\pi$ are defined by
\begin{eqnarray}
    \varphi(x)	& := & \phi(X(x)), 			\label{Def:varphi}\\
    \pi(x)      & := & (\det\gamma)^{1/2}(x)\, n^{\mu}(X(x))\,
					\partial_{\mu}\phi(X(x)),\label{Def:pi}
\end{eqnarray}
for all $x\in\Sigma$, and where $\gamma$ is defined in Eq.\
(\ref{Def:gamma}).

	We shall need certain Sobolev spaces of tensor-density fields on
$\Sigma$ (for more details see \cite{H+E}). The first step is to
introduce a fixed, auxiliary $C^r$ Riemannian metric $f_{kl}$ on
$\Sigma$. Let $I,J,K,\ldots$ denote multiple indices, and let $|I|$
be the number of simple indices within $I$. Let $f_{IJ}$ be the
abbreviation for the tensor product of $|I| = |J|$ copies of the
covariant tensors $f_{ij}$, and---similarly---$f^{IJ}$ denotes the
appropriate tensor product of contravariant fields $f^{ij}$. If
$T_I^J$ is a $C^r$ tensor-density field on $\Sigma$ of type
$(|I|,|J|,w)$ ($w$ is the weight), $T_{I|K}^{J}$ will denote the
tensor-density field obtained from $T_I^J$ by a $|K|$-fold covariant
derivative (covariant with respect to the auxiliary metric $f$).
Then the Sobolev space scalar product $(T,S)_f^s$ between two
tensor-density fields $T_I^J$ and $S_I^J$ is defined by
\begin{equation}
     (T,S)_f^s := \sum_{|M|=|N|=0}^s \int_{\Sigma} d^3x\,
          (\det f)^{1/2-w} f^{IK}f^{MN}f_{JL} T_{I|M}^J S_{K|N}^L
\end{equation}
where $s\leq r$. We denote the corresponding Sobolev space by
$H_{|I||J|w}^s(\Sigma)$, or simply $H_w^s(\Sigma)$ if no confusion
can result. Note that the topology of these spaces is independent of
the auxiliary metric $f$ provided that $\Sigma$ is compact (which we
are assuming).

	Using this notation, we can introduce the space $\Gamma^s_{\phi}$ of
Cauchy data:
\begin{equation}
     \Gamma^s_{\phi} := H_0^s(\Sigma) \times H_1^{s-1}(\Sigma)
\end{equation}
with $\varphi\in H_0^s(\Sigma)$ and $\pi\in H_1^{s-1}(\Sigma)$.  The
dynamics of the field $\phi$ as determined by Eq.\ (\ref{K-G})
defines maps between the spaces of Caucha data corresponding to
different embeddings.  The relevant facts about
these maps (which are more or less well-known; see
\cite{F+M}) are listed in Appendix 1, and culminate in the following theorem:
\begin{thm}
    Let $X_1$ and $X_2$ be two arbitrary $C^{r+1}$ embeddings with
$r\geq 4$. Then each pair $(\varphi_1,\pi_1)\in\Gamma^s_{\phi}$
(with $2\leq s\leq r-2$) defines a unique solution $\phi$ of the
field equation (\ref{K-G}) such that
\begin{enumerate}
	\item the Cauchy datum associated with the embedding $X_1$ is
the given pair $(\varphi_1,\pi_1)$;

\item the embedding $X_2$ gives a well-defined Cauchy datum
$(\varphi_2,\pi_2)$ that belongs to $\Gamma^s_{\phi}$;

\item the map $\rho_{X_{1}X_{2}}:\Gamma^s_{\phi} \rightarrow
\Gamma^s_{\phi}$ thus defined is an automorphism of the Sobolev
space $\Gamma^s_{\phi}$.

\end{enumerate}
\end{thm}
Note that we obtain a maximal classical solution in the sense that
the embedding $X_2$ can be chosen to map $\Sigma$ so that it passes
through any given point of $\cal M$.

	The theorem above implies that a differentiable solution is
obtained if $s$ is sufficiently large. Indeed, the famous Sobolev lemma
asserts that $H_w^s(\Sigma)\subset C_w^{s'}(\Sigma)$ if $ s > s' +
\frac{1}{2} {\text {dim}}\Sigma$. For example, $\phi$ will be $C^2$ if $s=4$.
The index $s$ can be taken as large as one wishes if $r$ is
sufficiently large. In particular, for $r=\infty$, one can take the
intersection $\Gamma^{\infty}_{\phi}$ of the Hilbert spaces $\Gamma
^s_{\phi}$, $s=4,5,\ldots$, to give the space of all pairs of
$C^{\infty}$-functions and densities that is equipped with the
structure of a countably Hilbert nuclear space (see \cite{G+V}).

\section{The extended phase space}
\label{Sec:extended_phase_space}
The theory of a scalar field on a curved background was rewritten in
the form of a parametrized system in \cite{I+K,kuch-prehled}.  In
this section, we shall reformulate a part of this work so that it
becomes compatible with the mathematical formalism of Fischer and
Marsden \cite{F+M}.  First however, the studies in
\cite{hyperspace,I+K} of the differential geometry of the space of
embeddings $\cal E$ must be extended to include certain bundles over
$\cal E$.

	The construction of a smooth differential structure on a space
of continuous maps between two finite-dimensional manifolds was
described as early as 1958 by Eells \cite{eells}, but we shall use
the method developed more recently in \cite{M+E+F}. Recall that, in
a pair of local charts
\begin{equation}
    (U,h){\text { of }}\Sigma{\text{ and }}(\bar{V},\bar{h})
		{\text{ of }} {\cal M} 			\label{charts}
\end{equation}
(where $X(U)\cap\bar{V}\neq\emptyset$), a given embedding
$X:\Sigma\rightarrow {\cal M}$ can be represented by the function
$\bar{h}\circ X\circ h^{-1}:h(U)\rightarrow\mathR^4$. We say that
$X$ belongs to the space $H^s(\Sigma,{\cal M})$ if these local
representatives are in the Sobolev space $H^s(h(U),\mathR^4)$ for
all such pairs of local charts. This notion can be shown to be
atlas-independent for $s>3/2$ (which means that $X$ is continuous
since, according to the Sobolev lemma, $X\in C^r({\cal N},{\cal M})$
if $s>r+{\text {dim}}({\cal N})/2$).

	In what follows, a major role is played by the tangent and
cotangent bundles to the infinite-dimensional manifold of
embeddings.  In the differential geometry of a finite-dimensional
manifold $\cal N$, a tangent vector $\tau$ at a point $p\in {\cal
N}$ can be defined in several different ways. One algebraic approach
is to view $\tau$ as a derivation at $p$ of the ring $C^\infty({\cal
N})$ of smooth functions on $\cal N$: {\em i.e.}, $\tau:
C^\infty({\cal N})\rightarrow\mathR$ is a linear map with the
property that if $f,g\in C^\infty({\cal N})$ then $\tau(fg)=
f(p)\tau(g)+g(p)\tau(f)$. A more geometrical approach is to define
$\tau$ as an equivalence class of local curves
$\sigma:(-\epsilon,\epsilon)\rightarrow {\cal N}$ where (i) $\epsilon>0$
(and can be $\sigma$-dependent); (ii) $\sigma(0)=p$; and (iii) two local
curves $\sigma_1$ and $\sigma_2$ are regarded as being equivalent if
their tangent vectors at $p\in {\cal N}$ are equal as computed in a
local coordinate system around $p$ (the equivalence classes are
independent of choice of coordinate system).

	In the finite-dimensional case, these two definitions can be
shown to be equivalent. However, the situation in infinite
dimensions is quite different since complicated
functional-analytical problems need to be resolved before the
algebraic definition can even be posed. Fortunately, the geometrical
definition of a tangent vector as an equivalence class of curves
still works well and, when applied to the case of interest, leads
naturally to the definition of the tangent space $T_X
H^s(\Sigma,{\cal M})$ to $H^s(\Sigma,{\cal M})$ at the embedding $X$
as the set of maps $V:\Sigma\rightarrow T{\cal M}$ with the property
that the image $V(x)$ of a point $x\in\Sigma$ is a tangent vector on
$\cal M$ at the point $X(x)$; {\em i.e.}, $V(x)\in T_{X(x)}{\cal
M}$; more formally, $V$ satisfies the relation $X=\chi\circ V$, where
$\chi:T{\cal M}\rightarrow {\cal M}$ is the bundle projection of the
tangent bundle $T{\cal M}$ of $\cal M$.

	This defining property of $V\in T_XH^s(\Sigma,{\cal M})$ can
be expressed in another way that will be useful later. Namely, we
recall that if $\rho:E\rightarrow{\cal M}$ is the projection map of
any fibre bundle $E$ over a manifold $\cal M$, and if $f:{\cal
N}\rightarrow {\cal M}$ is a map from another manifold $\cal N$ into
$\cal M$, then the pull-back bundle $f^*E$ over $\cal N$ is defined
as
\begin{equation}
	f^*E:=\{(x,e)\in {\cal N}\times E\mid f(x)=\rho(e)\},
\end{equation}
and a cross-section of this bundle is given by any map
$\psi:{\cal N}\rightarrow E$ such that $\rho(\psi(x))=f(x)$. It
follows therefore that a vector $V\in T_XH^s(\Sigma,{\cal M})$ can
be regarded as a cross-section of the bundle $X^*(T{\cal M})$.

	Note that the vector space $T_X H^s(\Sigma,{\cal M})$ can be
given an $H^s$-structure so that the function space
$H^s(\Sigma,{\cal M})$ becomes a Banach manifold modelled on the
Banach space $T_X H^s(\Sigma,{\cal M})$ (for example, via an
exponential map in $\cal M$). We shall assume from now on that this
has been done, and we shall consider only embeddings that lie in
$H^s(\Sigma,{\cal M})$, and (with the value $s$ of the Sobolev class
understood) denote the set of all such by ${\cal E}$; i.e., ${\cal
E}:= \text{Emb}_g(\Sigma,{\cal M})\bigcap H^s(\Sigma,{\cal M})$. It
can be shown that $\cal E$ is an open subset of $H^s(\Sigma,{\cal
M})$, and hence $\cal E$ is a Banach manifold with the same tangent
spaces and analogous manifold structure as that of $H^s(\Sigma,{\cal
M})$ itself.

	The above definition of a tangent vector leads immediately to
the definition of the tangent bundle $T{\cal E}$ of $\cal E$. More
generally, if ${}_S^RT{\cal M}$ is the tensor bundle of type
$(R,S)$ over $\cal M$ ($R$ times contravariant and $S$ times
covariant), we obtain an `$e$-tensor bundle' ${}_S^RT{\cal E}$
over $\cal E$ by defining an $e$-tensor at the point $X\in\cal E$ to
be a map $\psi:\Sigma\rightarrow {}_S^RT{\cal M}$ such that
$\psi(x)\in {}_S^R T_{X(x)}{\cal M}$; {\em i.e.}, $X =
\chi\circ\psi$, where $\chi$ now denotes the bundle projection of
${}_S^RT{\cal M}$. Equivalently, $\psi$ can be regarded as a
cross-section of the bundle $X^*({}_S^RT{\cal M})$ over $\Sigma$.

	We shall need an even more general $e$-tensor of the type defined in
\cite{hyperspace} which transforms as a tensor density of type $(r,s,w)$ with
respect to a coordinate change in $\Sigma$ around $x$, and as a
tensor of type $(R,S)$ with respect to a coordinate change in $\cal
M$ around the point $X(x)\in {\cal M}$. The precise definition of
such an $e$-tensor is that it is a cross-section $\psi$ of the
bundle ${}^{r,w}_sT\Sigma\otimes X^*({}_S^RT{\cal M})$ over
$\Sigma$, where ${}^{r,w}_sT\Sigma$ is the bundle of tensors over
$\Sigma$ that are $r$ times contravariant, $s$ times covariant, and
of tensor-density weight $w$. It follows that, for all $x\in\Sigma$,
we have $\psi(x)\in {}^{r,w}_sT_x\Sigma \otimes {}_S^RT_{X(x)}{\cal
M}$, and hence $\psi(x)$ can be represented by its components
$\psi_{l_1\ldots l_s\ \nu_1\ldots\nu_S}^{k_1\ldots k_r\
\mu_1\ldots\mu_R}(x)$ on $\Sigma$ defined with respect to an
appropriate pair of coordinate charts on $\Sigma$ and $\cal M$.

	There is another way of representing $e$-tensors which is
particularly useful for discussing tangent vectors to the collection
of all $e$-tensors of a certain type. Namely, recall that if $V$ and
$W$ are any pair of finite-dimensional vector spaces then there is a
canonical isomorphism of $V^*\otimes W$ with the space $L(V,W)$ of
linear maps from $V$ to $W$ in which the linear map $L_{\ell\otimes
w}$ associated with $\ell\otimes w\in V^*\otimes W$ is defined by
$L_{\ell\otimes w}(v):= \langle\ell,v\rangle w$ for all $v\in V$. In
particular, we note that ${}^{r,w}_sT_x\Sigma$ is the algebraic dual
of ${}^{s,-w}_rT_x\Sigma$, and hence it follows that if $\psi$ is an
$e$-tensor with $\psi(x)\in {}^{r,w}_sT_x\Sigma \otimes
{}_S^RT_{X(x)}{\cal M}$, then we can identify $\psi(x)$ as a linear
map $\psi(x): {}^{s,-w}_rT_x\Sigma \rightarrow {}^R_ST_{X(x)}{\cal
M}$. This can be summarised rather neatly by defining an $e$-tensor
to be a vector bundle map $\psi$ from ${}^{s,-w}_r T\Sigma$ to
${}^R_ST{\cal M}$, {\em i.e.}, the following diagram is commutative
\begin{equation}
\begin{array}{ccc}
{}^{s,-w}_rT\Sigma	& \stackrel\psi\longrightarrow &{}^R_ST{\cal M}\\
	\downarrow\rho &				   &  \downarrow\chi\\
 		\Sigma 	& \stackrel X  \longrightarrow &	{\cal M}
								\label{Def:qt}
\end{array}
\end{equation}
where $\rho$ and $\chi$ are the projection maps in the indicated
tensor bundles over $\Sigma$ and $\cal M$ respectively. Thus if
$\tau\in {}^{s,-w}_r T_x\Sigma$, we have $\psi(\tau)\in
{}^R_ST_{X(x)}{\cal M}$.

	We shall denote the bundle of such $e$-tensors by
${}^{R,r,w}_{S,s}T{\cal E}$. The linear space
${}^{R,r,w}_{S,s}T_X{\cal E}$ can be given a Sobolev structure with
an arbitrary Sobolev class (not necessarily the same as that of
$\cal E$). An easy, `covariant' method for doing so can be obtained
by using an auxiliary Riemannian metric on $\cal M$ in addition to
that on $\Sigma$. This enables covariant derivatives of $e$-tensors
to be defined (see \cite{hyperspace}).  However, a `coordinate
dependent' method is even easier to define, and is more general in
the sense that it can also be used for objects of higher rank, like
the elements of $T_{(X,P)}(T^*{\cal E})$ (see below).  This is based
on pairs of charts Eq.\ (\ref{charts}), which will associate a map
of $h(X^{-1}(X(U)\cap\bar{V}))\subset\mathR^3$ into $\mathR^m$ with
any $e$-tensor. One can then define a Sobolev scalar product by
patching together the integrands within each set
$h(X^{-1}(X(U)\cap\bar{V}))$ with the aid of a partition of unity
corresponding to a covering of $\Sigma$ by these sets. This scalar
product depends on the system of charts chosen, but the topology
does not and is equivalent to that obtained using the covariant
method.

	It is clear that the usual tensor operations like linear
combination, tensor product and contraction at a point $x\in\Sigma$
or $X(x)\in{\cal M}$, define corresponding operations on the
$e$-tensors (for details see \cite{hyperspace}). The result is an
$e$-tensor whose Sobolev class coincides with the lowest class
involved in the operation.

	There is one more operation of importance that we shall call
`pairing'. Let $\xi\in{}^{R,r,w}_{S,s}T_X{\cal E}$ and $\eta
\in{}^{S,s,1-w}_{R,r}T_X{\cal E}$. The pairing $\langle\xi,\eta\rangle$ is
defined by
\begin{equation}
     \langle\xi,\eta\rangle:=\int_{\Sigma}d^3x\,\xi(x)\cdot\eta(x)
\end{equation}
where $\xi(x)\cdot\eta(x)$ denotes the contractions at $x$ and
$X(x)$ such that all indices of $\xi$ are contracted with those of
$\eta$ in the order in which they appear. Then $\xi(x)\cdot\eta(x)$
is a scalar density on $\Sigma$, and hence $\langle\xi,\eta\rangle$
is a coordinate independent real number.

	A particularly important example of a bundle of $e$-tensors is
the tangent bundle $T{\cal E}:={}^{1,0,0}_{0,0}T{\cal E}$ whose
points $\xi$ are pairs $(X,V)$ where $X\in\cal E$, and $V$ is a
$T{\cal M}$-valued function on $\Sigma$ with $V(x)\in T_{X(x)}{\cal
M}$; equivalently, $V$ is a cross-section of $X^*(T{\cal M})$. Even
more important for our purposes is the cotangent bundle $T^*{\cal
E}$. To define cotangent vectors we have to identify $T_X^*{\cal E}$
with a particular topological vector space of real-valued, linear
functions on $T_X\cal E$.  The appropriate choice for our purposes
is the space ${}^{0,0,1}_{1,0}T_X{\cal E}$ of Sobolev class $s'$,
the linear operation being the pairing (the class parameter $s'$
must satisfy the condition $s'\leq s$, where $s$ is the class
parameter of $\cal E$, but can otherwise be arbitrary). Thus, $P\in
T^*_X{\cal E}:={}^{0,0,1}_{1,0}T_X{\cal E}$ is a cross-section of
the bundle ${\cal D}^1\Sigma\otimes X^*(T^*{\cal M})$ where, in
general, ${\cal D}^w\Sigma$ is shorthand for the real-line bundle
${}^{0,w}_0T\Sigma$ of scalar densities on $\Sigma$ of weight $w$.
Thus $P(x)\in {\cal D}^1_x\Sigma\otimes T^*_{X(x)}{\cal M}$;
equivalently, $P$ is a bundle map from ${\cal D}^{-1}\Sigma$ to
$T^*{\cal M}$ that `covers' $X$ in the sense that $P(\tau)\in
T^*_{X(x)}{\cal M}$ for all $\tau\in{\cal D}^{-1}_x\Sigma$.

  With respect to a local coordinate chart $y^\mu$, $\mu=0,1,2,3$,
on $\cal M$, the pair $(X,P)\in T^*{\cal E}$ is represented by local
functions $(X^\mu,P_\nu)$ on $\Sigma$, where $X^\mu$ is defined by
\begin{equation}
		X^\mu(x):=y^\mu(X(x)),
\end{equation}
and where the four scalar-density functions $P_{\nu}$,
$\nu=0,1,2,3$, on $\Sigma$ are defined by the expansion
\begin{equation}
			P(x)=P_\nu(x)\,dy^\nu\vert_{X(x)}
\end{equation}
using the differentials $dy^\nu$ associated with the local
coordinate system $y^\nu$ on $\cal M$ (as usual, summation over
repeated indices is understood) and using a local coordinate system
on $\Sigma$ to locally-trivialise the line bundle ${\cal
D}^{-1}\Sigma$.  In terms of these local functions, the pairing
operation is
\begin{equation}
   \langle(X,P),(X,V)\rangle =\int_{\Sigma}d^3x\,P_{\mu}(x)V^{\mu}(x)
\end{equation}
where the functions $V^{\mu}$ on $\Sigma$ are defined by
the equation
\begin{equation}
	V(x)=V^\mu(x)\left({\partial\over\partial y^\mu}\right)_{X(x)}.
\end{equation}

	The $e$-tensor bundles can be given a manifold structure based
on that of $\cal E$. A point in such a bundle
${}^{R,r,w}_{S,s}T{\cal E}$ is represented by a pair $(X,\psi)$
where $X\in\cal E$, and $\psi\in {}^{R,r,w}_{S,s}T_X{\cal E}$ is a
cross-section of the bundle ${}^{r,w}_sT\Sigma\otimes
X^*({}_S^RT{\cal M})$ or---better for our purposes---the pair
$(X,\psi)$ fits into the commutative diagram Eq.\ (\ref{Def:qt}).  A
tangent vector is defined as an equivalence classes of curves
$t\mapsto(X_t,\psi_t)$, and it is clear from this geometrically that
a tangent vector at $(X,\psi)$ consists of a pair of objects $(V,W)$
where (i) $V\in T_X{\cal E}$ ({\em i.e.}, $V(x)\in T_{X(x)}{\cal
M}$); (ii) $W$ is a bundle map from ${}^{s,-w}_rT\Sigma$ to
$T({}^R_ST{\cal M})$ such that, for all $\tau\in
{}^{s,-w}_rT_x\Sigma$, we have $W(\tau)\in
T_{\psi(\tau)}({}^R_ST{\cal M})$; and (iii) $\chi_*(W(\tau))=V(x)$ where
$\chi_*: T({}^R_ST{\cal M})\rightarrow {}T{\cal M}$ is induced from
the bundle projector $\chi:{}^R_ST{\cal M}\rightarrow{\cal M}$ in
Eq.\ (\ref{Def:qt}). Note that $V$ and $W$ both take their values in
vector spaces, and hence the collection of all such pairs $(V,W)$
can be given an appropriate Sobolev structure to become Banach
spaces. By this means, the bundle of $e$-tensors over $\cal E$
becomes a Banach manifold.

		Now, in general, if $E$ is any vector bundle over $\cal M$,
the tangent space $T_pE$ splits into a direct sum $E_{\pi(p)}\oplus
T_{\pi(p)}{\cal M}$ where $E_{\pi(p)}$ denotes the fiber of $E$ over
the point $\pi(p)\in \cal M$.  However, in the absence of any
connection on $E$ there is no natural way of performing such a
split. Note that in our case, where $E={}^R_ST{\cal M}$, this
(non-canonical) split means that the vector $W(\tau)\in
T_{\psi(\tau)}({}^R_ST{\cal M})$ (with
$\tau\in{}^{s,-w}_rT_x\Sigma$) can be written as a sum of an element
of ${}^R_ST_{X(x)}{\cal M}$ (the analogue of $E_{\pi(p)}$) and an
element (in fact, $V(x)$) of $T_{X(x)}{\cal M}$. This means that,
using a local coordinate system on $\cal M$, the vector $W(\tau)$ can
be written as a collection of numbers associated with the point
$X(x)\in {\cal M}$, namely the components $V^\mu(x)$ of the vector
$V(x)$, and the components of the space-time object in
${}^R_ST_{X(x)}{\cal M}$, which we shall write as
$W^{\mu_1\ldots\mu_R}_{\nu_1\ldots\nu_S}(\tau)$.

 	A particularly simple example is the tangent bundle $T(T^*\cal
E)$.  The bundle $T^*\cal E$ consists of pairs $(X,P)$ where
$X\in\cal E$, and where $P$ is a bundle map $P:{\cal D}^{-1}\Sigma
\rightarrow T^*\cal M$ that covers $X:\Sigma\rightarrow {\cal M}$,
{\em i.e.}, $P(\tau)\in T^*_{X(x)}{\cal M}$ for all $\tau\in{\cal
D}^{-1}_x\Sigma$. Then, according to the discussion above, a tangent
vector to $T^*{\cal E}$ at the point $(X,P)$ consists of a pair
$(V,W)$ where $V\in T_X{\cal E}$ and where $W:{\cal
D}^{-1}\Sigma\rightarrow T(T^*{\cal M})$ satisfies $W(\tau)\in
T_{P(\tau)}(T^*{\cal M})$ with $\pi_*(W(\tau))= V(x)$ for all
$\tau\in {\cal D}^{-1}_x\Sigma$, where $\pi:T^*{\cal M}\rightarrow
{\cal M}$ is the bundle projection.

 	The problem occasioned by the non-canonical split can be seen by
looking at the situation using a local coordinate system.  The
element $(X,P)\in T^*{\cal E}$ is represented by the local functions
$(X^{\mu},P_{\nu})$ on $\Sigma$, and a transformation of coordinates
in $\cal M$ from $y^\mu$ to $y^{'\alpha}$ leads to new functions
$(X^{'\alpha},P'_\beta)$ satisfying
\begin{eqnarray}
     X^{'\alpha}(x) & = & y^{'\alpha}(X(x)) \label{tran1} \\
     P'_{\beta}(x) & = & J^\nu_{\beta'}(X(x))\,P_\nu(x), \label{tran2}
\end{eqnarray}
where $J^{\nu}_{\beta'}$ denotes the matrix $\partial y^{\mu}/
\partial y^{'\beta}$. As emphasised earlier, a tangent vector in
infinite-dimensional differential geometry is defined as an
equivalence class of curves and, with respect to the pair of charts
Eq.\ (\ref{charts}), a curve in $T^*\cal E$ is represented by
functions $(t,x)\mapsto\left(X^{\mu}(x,t),P_{\nu}(x,t)\right)$.
Hence the tangent vector to this curve is represented by the
functions $\dot{X}^\mu$ and $\dot{P}_\nu$ on $\Sigma$ where
\begin{equation}
\dot{X}^\mu(x):=
	\left.{\partial X^\mu(x,t)\over\partial t}\right|_{t=0},\qquad
\dot{P}_\nu(x):=
	\left.{\partial P_\nu(x,t)\over\partial t}\right|_{t=0}.
\end{equation}
Note that $P_{\mu}(x,t)$ are components with respect to
coordinates at the point $X(x,t)$ of $\cal M$ while $P_{\mu}(x,t+dt)$ are
those at another point $X(x,t+dt)$ of $\cal M$.

	Differentiating formula Eq.\ (\ref{tran1}) along the curve with
respect to $t$, we obtain
\begin{eqnarray}
	\dot{X}^{'\alpha}(x)&:=&
	\left.{\partial X^{'\alpha}(x,t)\over\partial t}\right|_{t=0}
  = \left.{\partial y^{'\alpha}(X(x,t))\over\partial t}\right|_{t=0}
	= {\partial y^{'\alpha}\over\partial y^\mu}(X(x))\,
	\left.{\partial y^\mu(X(x,t))\over\partial t}\right|_{t=0} \nonumber \\
	&=& J^{\alpha'}_\mu(X(x))\, \dot{X}^\mu(x),
\end{eqnarray}
which shows that, as expected, the derivatives $\dot{X}^\mu(x)$
transform on $\cal M$ in a tensorial way. On the other hand, the
analogous calculations for Eq.\ (\ref{tran2}) yield
\begin{equation}
	\dot{P}^{'}_\beta(x)=J^\nu_{\beta'\mu}(X(x))
	\dot{X}^\mu(x)P_\nu(x)+J^\nu_{\beta'}(X(x))\, \dot{P}_\nu(x)
\end{equation}
where $J^\nu_{\beta'\mu}$ denotes the derivative of the matrix
$J^\nu_{\beta'}$ with respect to the original coordinates
$y^\mu$.  This shows that the four quantities
$\dot{P}^{'}_\beta(x)$ cannot be regarded as the components of
any tensorial object on $\cal M$. In general, a tangent vector to
$(X,P)\in T^*{\cal E}$ can be represented by the functions
$(V^\mu,W_\nu)$ on $\Sigma$ which transform under a change of
coordinates on $\cal M$ as
\begin{equation}
\left(\begin{array}{c}
			V^{'\alpha}(x)\\
			W^{'}_\beta(x)
	  \end{array}\right)=
		\left(\begin{array}{ll}
				J^{\alpha'}_\mu(X(x)) 	& 	0	\\
				J^\rho_{\beta'\mu}(X(x))
					P_\rho(x) & J^\nu_{\beta'}(X(x))
			  \end{array}\right)
							\left(\begin{array}{c}
								V^\mu(x)\\
								W_\nu(x)
					\end{array}\right). \label{transvect}
\end{equation}

	We shall often need to work with the cotangent vectors from
$T^*(T^*{\cal E})$.  Let us describe their most important
properties. As discussed above, a vector in $T_{(X,P)}(T^*{\cal E})$
is a pair $(V,W)$ where $V\in T_X{\cal E}$, and $W:{\cal
D}^{-1}\Sigma\rightarrow T(T^*{\cal M})$ is such that
$\pi_*(W(\tau))=V(x)\in T_{X(x)}{\cal M}$ for all $\tau\in {\cal
D}^{-1}_x\Sigma$, where $\pi:T^*{\cal M}\rightarrow {\cal M}$ is the
bundle projection. By definition, the kernel of the map
$\pi_*:T(T^*{\cal M})\rightarrow T{\cal M}$ is the set of vertical
vectors in $T(T^*{\cal M})$, and at each $k\in T^*{\cal M}$ there is
a natural isomorphism \(\iota\) of $V_k(T^*{\cal M})$ (the vertical
tangent vectors at $k$) with $T^*_{\pi(k)}{\cal M}$. In fact, at
each $k\in T^*{\cal M}$, the map $\pi_*$ fits into the short exact
sequence
\begin{equation}
0	\longrightarrow T^*_{\pi(k)}{\cal M}\stackrel\iota
 	\longrightarrow T_k(T^*{\cal M})\stackrel{\pi_*}
	\longrightarrow T_{\pi(k)}{\cal M}\longrightarrow  0.\label{SEQ1}
\end{equation}
The dual of this sequence is the short exact sequence
\begin{equation}
0\longrightarrow T^*_{\pi(k)}{\cal M}\stackrel{\pi_*^\dagger}\longrightarrow
T_k^*(T^*{\cal M})\stackrel{\iota^\dagger}\longrightarrow
T^{**}_{\pi(k)}{\cal M}\longrightarrow 0			\label{SEQ2}
\end{equation}
where a `$\dagger$' superscript denotes the adjoint of the linear
map to which it is attached. Note that, since $\cal M$ is
finite-dimensional, the third term in Eq.\ (\ref{SEQ2}), ({\em i.e.},
$T^{**}_{\pi(k)}{\cal M})$ is (non-canonically) isomorphic to
$T_{\pi(k)}{\cal M}$.

	Using the ideas above, it is natural to define an element of
$T^*_{(X,P)}(T^*{\cal E})$ as a pair $(A,B)$ where
$B:\Sigma\rightarrow T^{**}{\cal M}\simeq T{\cal M}$, and where the
bundle map $A:{\cal D}^{-1}\Sigma\rightarrow T^*(T^*{\cal M})$
satisfies $A(\tau)\in T^*_{P(\tau)}(T^*{\cal M})$ with
$\iota^\dagger(A(\tau))=B(x)$ for all $\tau\in {\cal
D}^{-1}_x\Sigma$.  In local coordinates on $\cal M$, an element
$(A,B)\in T_{(X,P)}^*(T^*{\cal E})$ can be represented by a set of
local functions $(A_\mu,B^\nu)$ on $\Sigma$ (each $A_\mu$ is
actually a scalar density on $\Sigma$; $B^\nu$ are genuine scalar
functions). The pairing of such an object with a vector $(V,W)$ in
$T_{(X,P)}(T^*{\cal E})$ is defined by
\begin{equation}
     \langle (A,B),(V,W)\rangle
	:=\int_\Sigma d^3x\,(A_\mu(x)V^\mu(x)+ B^\mu(x) W_\mu(x)).
\end{equation}
The condition that the result be independent of coordinates on $\cal
M$ determines the transformation of $(A_\mu(x),B^{\nu}(x))$ to
be
\begin{equation}
\left(\begin{array}{c}
          A^{'}_\alpha(x)\\
          B^{'\beta}(x)
      \end{array} \right) =
     	\left(\begin{array}{ll}
          J_{\alpha'}^\mu & -J^\rho_{\alpha'\nu}(X(x))\,
		P_\rho(x)\\
          0 & J^{\beta'}_\nu(X(x))
     		  \end{array} \right)
     				\left(\begin{array}{l}
          				A_\mu(x) \\
          				B^\nu(x)
     				\end{array} \right).  \label{transcov}
\end{equation}
Thus $V^\mu(x)$ and $B^\mu(x)$ represent tensorial objects on
$\cal M$, but $W_\mu(x)$ and $A_\mu(x)$ do not.

	A key observation is the following. If the transformation law
Eq.\ (\ref{transvect}) is compared with Eq.\ (\ref{transcov}),
it is clear that the quantity $(G^{\mu},-F_{\mu})$ formed from the
`components' of a covector on $T^*{\cal E}$ transforms as a tangent
vector on $T^*{\cal E}$.  Thus we have a map $J_{\cal E} : T^*
(T^*{\cal E})\rightarrow T(T^*{\cal E})$
defined in local coordinates on the component functions by
\begin{equation}
     J_{\cal E}(A_\alpha,B^\nu) = (B^\mu,-A_\alpha).	\label{Def:J}
\end{equation}
We can choose the Sobolev classes so that the classes of the
functions $V^\alpha$, $X^\alpha$ and $B^\alpha$ coincide, and so do
those of $P_\beta$, $W_\beta$ and $A_\beta$. Then $J_{\cal E}$ is a
Sobolev space isomorphism.

	More abstractly, we recall that if $Q$ is any
finite-dimensional manifold there is a canonical isomorphism $j:
T^*_p(T^*Q)\rightarrow T_p(T^*Q)$ defined on any cotangent vector
$\ell$ at the point $p$ in $T^*Q$ by
\begin{equation}
	\omega(j(\ell),v)=\langle\ell,v\rangle_p, {\text{ for all }}
					v\in T_p(T^*Q)	\label{Def:j}
\end{equation}
where $\omega$ is the canonical two-form on the cotangent bundle
$T^*Q$. In the context of the embedding space, if $(A,B)\in
T^*_{(X,P)}(T^*{\cal E})$ then $A(\tau)\in T^*_{P(\tau)}(T^*{\cal
M})$, and hence we can use the isomorphism $j:T^*_{P(\tau)}(T^*{\cal
M})\rightarrow T_{P(\tau)}(T^*{\cal M})$ to define a map
$J:T^*_{(X,P)}(T^*{\cal E})\rightarrow T_{(X,P)}(T{\cal E})$ by
requiring that $J(A,B)(\tau):=j(A(\tau))$. This is the coordinate-free
definition of the object given in Eq.\ (\ref{Def:J}).

	The phase space $\Gamma_\phi$ of the scalar field $\phi$ was
introduced in section \ref{Sec:dynamics}. We note now that
$\Gamma_{\phi}$ can be considered as the cotangent bundle $T^*{\cal
Q}$, where $\cal Q$ is the space $H_0^{\infty}$ of all
$C^{\infty}$-scalar fields on $\Sigma$.  As $\cal Q$ is a linear
space, there is a natural identification between the spaces
$T^*{\cal Q}$ and ${\cal Q}\times T_{\varphi}^*{\cal Q}$ for any
$\varphi\in{\cal Q}$. Also, $T_{\varphi}^*{\cal Q} = H_1^{\infty}$
is itself a linear space, and so there is an identification between
the spaces $T_{(\varphi ,\pi)}(T^*{\cal Q})$ and $T^*{\cal Q}\simeq
H_0^{\infty} \times H_1^{\infty}$.  Similarly, $T_{(\varphi,\pi)}^*
(T^*{\cal Q})\simeq H_1^{\infty} \times H_0^{\infty}$. More
precisely, if $(\xi,\eta)\in T_{(\varphi,\pi)}(T^*{\cal Q})$ and
$(f,h)\in H_1^{\infty} \times H_0^{\infty}$, then the pairing $(f,h)
: T_{(\varphi,\pi)}(T^*{\cal Q}) \rightarrow\mathR$ is defined by
\begin{equation}
     \langle(f,h), (\xi,\eta)\rangle:= \int_\Sigma d^3 x\,(f\xi+h\eta).
\end{equation}

	The extended phase space $\Gamma$ of the parametrized scalar field of
\cite{I+K} is defined as $\Gamma := \Gamma_{\phi}\times T^*{\cal E}$.
Hence $\Gamma\simeq T^*({\cal Q}\times{\cal E})$, so that the extended
phase space is again a cotangent bundle.  In the context of the full
space $\Gamma$, a tangent vector from $T_{(\varphi,\pi,X,P)}\Gamma$
can be specified by its `components' $(\Phi,\Pi,V,W)$, where
$(V,W)\in T_{(X,P)}(T^*{\cal E})$ and $(\Phi,\Pi)\in T_{(\varphi,\pi
)}(T^*{\cal Q})$. Similarly, a cotangent vector from
$T_{(\varphi,\pi,X,P)}^*\Gamma$ can be specified by
$(A_{\varphi},A_{\pi},A,B)$, where $(A,B) \in T_{(X,P)}^*(T^*{\cal
E})$ and $(A_{\varphi},A_{\pi})\in T^*(T^*{\cal Q})$. The natural
pairing is
\begin{equation}
 \langle(A_{\varphi},A_{\pi},A,B),(\Phi,\Pi,V,W)\rangle :=
     \int_\Sigma d^3 x\,(A_{\varphi}\Phi+A_{\pi}\Pi+A_\mu V^{\mu}+
		     B^\mu W_{\mu}).
\end{equation}

	There is an isomorphism
$J:T_{(\varphi,\pi,X,P)}^*\Gamma\rightarrow
T_{(\varphi,\pi,X,P)}\Gamma$ (if the Sobolev classes are chosen to
match each other) given by
\begin{equation}
     J(A_{\varphi},A_{\pi},A_X,A_P) := (A_{\pi},-A_{\varphi},A_P,-A_X).
\end{equation}
Using this isomorphism, a symplectic structure
$\Omega$ on $\Gamma$ can be defined as follows. Let $v_1$ and $v_2$ be
two vectors in $T_{(\varphi,\pi,X,P)}\Gamma$. Then
\begin{equation}
     \Omega(v_1,v_2) := -\langle J^{-1}v_1, v_2\rangle	\label{Def:Omega}
\end{equation}
or, in `component' form,
\begin{equation}
     \Omega((\Phi_1,\Pi_1,V_1,W_1),(\Phi_2,\Pi_2,V_2,W_2)) =
     \int_\Sigma d^3 x\,(\Pi_1\Phi_2-\Phi_1\Pi_2 + W_{1\mu}V_2^{\mu} -
     				V_1^{\mu}W_{2\mu}).
\end{equation}
It follows at once that (i) $\Omega(v_1,v_2 ) = -\Omega(v_2,v_1)$;
(ii) $\Omega$ is weakly non-degenerate (see \cite{C+M}); and (iii)
$\Omega$ is not only closed but also exact.

	As $\Omega$ is only a weak symplectic form, not every
differentiable function on $\Gamma$ will have a Hamiltonian vector field.
The class of functions that do can be characterized as follows.
If $F:\Gamma\rightarrow\mathR$, we say that $F$ has a gradient if the
following two conditions are satisfied:
\begin{enumerate}
     \item the Fr\'{e}chet derivative, ${\text{D}}F|_{(\varphi,\pi,X,P)}:
			T_{(\varphi,\pi,X,P)}\Gamma\rightarrow\mathR$ is a
			bounded linear map;
     \item there exists ${\text{grad}}\,F\in
		T_{(\varphi,\pi,X,P)}^*\Gamma$ such that
          $\langle{\text{grad}}\,F,v\rangle
		 = {\text{D}}F|_{(\varphi,\pi,X,P)}(v)$ for all
		$v\in T_{(\varphi,\pi,X,P)}\Gamma$.
          The `components' of this gradient will be denoted by
          the collection of functions
			$({\text{grad}}_{\varphi}F, {\text{grad}}_{\pi}F,
	          	({\text{grad}}_X F)_{\mu}, ({\text{grad}}_P F)^{\nu})$.

\end{enumerate}
Condition 2 means that ${\text{D}}F$ must be regular (no
distributions are accepted!), and hence we have to work with smeared
objects.  This will not lead to any real loss of generality. The
quantity ${\text{grad}} F$ is calculated from ${\text{D}}F$ as usual by
integration by parts (if $F$ contains derivatives).

	For a differentiable function with a gradient, we can define an
associated `Hamiltonian vector field'.  Specifically, if $F$ is such
a function, then $\xi_F\in T_{(\varphi,\pi,X,P)}\Gamma$ is defined
by the relation
\begin{equation}
    \langle{\text{grad}}\,F, v\rangle=\Omega(v,\xi_F),{\text{ for all }} v\in
		T_{(\varphi,\pi,X,P)}\Gamma.
\end{equation}
Hence, because of Eq.\ (\ref{Def:Omega}), we see that
$\langle{\text{grad}}\,F, v\rangle = \langle J^{-1}\xi_F, v\rangle$
for all $v$, and so $\xi_F = J({\text{grad}}\,F)$.

	Finally, the Poisson bracket of a pair of differentiable
functions  $F$ and $G$ is defined as $\{F,G\} :=
-\Omega(\xi_F,\xi_G)$, and we see immediately that
\begin{equation}
     \{F,G\} = \langle{\text{grad}}\,F,\xi_G\rangle. \label{poisson}
\end{equation}
This Poisson bracket is antisymmetric and, since $\Omega$ is closed, it
satisfies the Jacobi identity.

\section{The constraint manifold}
\label{Sec:constraint_manifold}
The parametrized scalar field theory possesses a set of constraints
${\cal H}_\mu=0$ on the Cauchy data $(\varphi,\pi,X,P)$. These constraints
are contained in the map
\begin{eqnarray}
     			C:\Gamma& \rightarrow 	& T^*{\cal E}, \\
     (\varphi,\pi,X,P) 	&	\mapsto 	& (X,{\cal H}(\varphi,\pi,X,P))
\end{eqnarray}
where ${\cal H}(\varphi,\pi,X,P)\in T^*_X{\cal E}$, {\em i.e.},
${\cal H}(\varphi,\pi,X,P)(x)\in T^*_{X(x)}({\cal M})$ for all
$x\in\Sigma$, so that ${\cal H}(\varphi,\pi,X,P)(x)={\cal
H}(\varphi,\pi,X,P)_\mu(x)\,dy^\mu|_{X(x)}$ in a coordinate system
$y^\mu$ on $\cal M$.  The specific form of the constraints ${\cal
H}_{\mu}(x)$ is given in
\cite{I+K} as
\begin{eqnarray}
  {\cal H}_{\mu} 		&=& P_{\mu}+{\cal H}_{\mu}^{\phi}, \\
								\label{cst1}
  {\cal H}_{\mu}^{\phi} & = & -{\cal H}_{\bot}^{\phi}\,n_{\mu} +
				{\cal H}_k^{\phi}\,X_{\mu}^k,	\label{cst2}
\end{eqnarray}
where $n_\mu(x)$ is the unit normal space-time vector at
$X(x)\in\cal M$, and where $X_{\mu}^k(x) :=
g_{\mu\nu}(x)\,\gamma^{kl}(X(x))\,X,_{l}^{\nu}(x)$.  The components
${\cal H}^\phi_\mu(\varphi,\pi,X)$ can be projected normal, and
tangential, to the hypersurface $X(\Sigma)$ to give
\begin{eqnarray}
  {\cal H}_{\bot}^{\phi}& = & \frac{1}{2}(\det\gamma)^{1/2}
	\left(\frac{\pi^2}{\det\gamma} + \gamma^{kl}\,\varphi_k\varphi_l
		 + m^2\,\varphi^2 \right), 		\\      \label{cst3}
  {\cal H}_k^{\phi} 	& = & \pi\,\varphi,_{k}. 		\label{cst4}
\end{eqnarray}

	The goal of this section is to explore the pre-symplectic
structure of the manifold $\tilde\Gamma$ of solutions to these
constraints, {\em i.e.}, $\tilde{\Gamma} := \{(\varphi,\pi,X,P)\in
\Gamma\mid C(\varphi,\pi,X,P) = (X,0)\}$. In particular, we have the
following theorem
\begin{thm}
     Let the Sobolev class of all spaces involved be $\infty$.
Then, $\tilde{\Gamma}$ is a $C^{\infty}$-submanifold of $\Gamma$ in
a neighbourhood of any of its points. This constraint manifold
$\tilde{\Gamma}$ is given by $\tilde{\Gamma} =
\tilde{C}(\Gamma_{\phi} \times {\cal E})$, where
$\tilde{C}:\Gamma_{\phi}\times{\cal E}\rightarrow\Gamma$ is defined
by $\tilde{C}(\varphi,\pi,X):= (\varphi,\pi,X,-{\cal
H}^{\phi}(\varphi,\pi,X))$.
\end{thm}

{\bf Proof} The proof is similar to that in \cite{F+M}, but requires
less sophisticated functional analysis because our constraints are
available in an explicit form; in particular---unlike the case in
\cite{F+M}---we do not need to use the Fredholm alternative theorem.

	We start by assuming that $T^*\Gamma$ is $C^{\infty}$, and
postulate that the Sobolev class of $\Pi$ and $W$ is $s-1$, where $s$
is the class of $X$, $\Phi$ and $V$. Consider the map
${\text{D}}C|_{(\varphi,\pi,X,P)} : T_{(\varphi,\pi,X,P)}\Gamma
\rightarrow T_{{\cal H}(\varphi,\pi,X,P)}(T^*{\cal E})$.
To use the implicit function theorem, this map must be a surjection
with a splitting kernel (see, e.g. \cite{Lang}). However,
${\text{D}}C|_{(\varphi,\pi,X,P)}$ is trivially surjective. Moreover,
its kernel in $T_{\tilde{C}(X,\varphi,\pi)}\Gamma$ is given by
${\text{D}}\tilde{C}|_{(\varphi,\pi,X)}
\left(T_{(\varphi,\pi,X)}(\Gamma_{\phi}\times{\cal E})\right)$.  We
shall now show that ${\text{D}}C|_{(\varphi,\pi,X)}$ is injective with
a splitting image.

	The map
${\text{D}}\tilde{C}|_{(\varphi,\pi,X)}:
     T_{(\varphi,\pi ,X)}(\Gamma_{\phi}\times{\cal E})\rightarrow
     T_{\tilde{C}(\varphi,\pi,X)}\Gamma$
between the Sobolev spaces
$T_{(\varphi,\pi ,X)}(\Gamma_{\phi}\times {\cal E}) \simeq
\Gamma^s_{\phi}\times H^s_0 $ and
$T_{\tilde{C}(\varphi,\pi ,X)}\Gamma \simeq \Gamma^s_{\phi}\times H^s_0
 \times H^{s-1}_1$ has a derivative given by
\begin{equation}
     {\text{D}}\tilde{C}|_{(\varphi,\pi,X)}(\Phi,\Pi,V)
     = (\Phi,\Pi,V, -{\text{D}}{\cal H}^{\phi}|_{(\varphi,\pi,X)}(\Phi,\Pi,V)),
								\label{DC}
\end{equation}
where
\begin{equation}
     {\text{D}}{\cal
H}^{\phi}|_{(\varphi,\pi,X)}(\Phi,\Pi,V)=
     {\text{D}}_{\varphi}{\cal H}^{\phi}|_{(\varphi,\pi,X)}(\Phi) +
     {\text{D}}_{\pi}{\cal H}^{\phi}|_{(\varphi,\pi,X)}(\Pi) +
     {\text{D}}_X {\cal H}^{\phi}|_{(\varphi,\pi,X)}(V). \label{DHphi}
\end{equation}
Using the results of \cite{I+K} and Eqs.\ (\ref{cst1}--\ref{cst4}),
we find after some calculation that
\begin{eqnarray}
     {\text{D}}_{\varphi}{\cal H}^{\phi}_{\mu}|_{(\varphi,\pi,X)}(\Phi)
     & = & (\det\gamma)^{1/2}(L^k_{\mu}\Phi_{\| k} - n_{\mu} m^2
     \varphi\Phi), \label{H1} \\
     {\text{D}}_{\pi}{\cal H}^{\phi}_{\mu}|_{(\varphi,\pi,X)}(\Pi) & = &
     L^{\bot}_{\mu}\Pi , \label{H2} \\
     {\text{D}}_X {\cal H}^{\phi}_{\mu}|_{(\varphi,\pi,X)}(V) & = &
     \frac{1}{2}\,(\det
     \gamma)^{1/2} K^k_{\mu\nu}V^{\nu}_{\| k} + {\cal H}^{\phi}_{\kappa}
     \Gamma^{\kappa}_{\mu\nu} V^{\nu} , \label{H3}
\end{eqnarray}
where
\begin{eqnarray}
     L^k_{\mu} & = & -\gamma^{kl}\varphi_{\| l}\,n_{\mu} +
     \frac{\pi}{(\det \gamma)^{1/2}}X^k_{\mu} , \label{L1} \\
     L^{\bot}_{\mu} & = & - \frac{\pi}{(\det \gamma)^{1/2}}\,n_{\mu} +
     \varphi_{\| k}\,X^k_{\mu} , \label{L2} \\
     K^m_{\mu\nu} & = & \left( \frac{\pi^2}{\det \gamma} - \gamma^{kl}
     \varphi_{\| k}\varphi_{\| l} - m^2 \varphi^2 \right)
     n_{\mu}X^m_{\nu} \nonumber \\
     & & + \left( \frac{\pi^2}{\det \gamma} + \gamma^{kl}
     \varphi_{\| k}\varphi_{\| l} + m^2 \varphi^2 \right)X^m_{\mu}n_{\nu} +
     2\,\varphi_{\| k}\,\varphi_{\| l}\,
     \gamma^{km}n_{\mu}X^k_{\nu}  \nonumber \\
     & & -2\frac{\pi}{(\det \gamma)^{1/2}}\,\varphi_{\| k}\,(\gamma^{km}
     n_{\mu}n_{\nu} + X^k_{\nu}X^m_{\mu}) , \label{K}
\end{eqnarray}
and `$\|$' denotes the bi-covariant derivative, for example
$ Y^{\mu}_{k\| l} = Y^{\mu}_{k,l} +
\Gamma^{\mu}_{\rho\sigma}Y^{\rho}_k X^{\sigma}_l - \gamma^m_{kl}Y^{\mu}_m$,
where $\Gamma^{\mu}_{\rho\sigma}$ is the Christoffel symbol of the
metric $g_{\mu\nu}$ on $M$, and $\gamma^m_{kl}$ is the Christoffel
symbol of the pull-back of $g$ to $\Sigma$ by $X$. By inspection,
these formulas imply the crucial result that the map ${\text{D}}{\cal
H}^{\phi}|_{(\varphi,\pi,X)}:H^s_0\times
\Gamma^s_{\phi}\rightarrow H^{s-1}_1$ is continuous and bounded
(recall that $\varphi$, $\pi$ and $X$ are $C^{\infty}$). Hence, for
all $(\Phi,\Pi,V)\in \Gamma^s_{\phi}\times H^s_0$, the map
${\text{D}}\tilde{C}|_{(\varphi,\pi,X)}$ is continuous and bounded,
and---by inspection---injective. It follows from the closed graph
theorem that the image
${\text{D}}\tilde{C}|_{(\varphi,\pi,X)}(\Gamma^s_{\phi}\times H^s_0)$
is therefore closed in $\Gamma^s_{\phi}\times H^s_0\times
H^{s-1}_1$.  However, a closed subspace of a Hilbert space splits,
which proves the theorem. \hfill {\bf QED}

	By abuse of language, we will often call $\Gamma_{\phi}\times
\cal E$ the `constraint manifold'.

	For each fixed value of $x$ and $\mu$, the quantity ${\cal
H}_{\mu}(x)$ can be viewed as a function on the phase space, but it
has no gradient; in particular, the Poisson bracket of a pair of
such functions is not well-defined. Constraint functions with
gradients can be constructed by `smearing'.  Specifically, if $N\in
T_X{\cal E}$ ({\em i.e.}, $N(x)\in T_{X(x)}{\cal M}$), then ${\cal
H}_N$ is defined as
\begin{equation}
     {\cal H}_N := \int_{\Sigma}d^3x\, N^{\mu}(x){\cal H}_{\mu}(x).
\end{equation}
It is clear that the equations ${\cal H}_N = 0$ for all $N\in
T_X\cal E$ are equivalent to ${\cal H}_{\mu}(x) = 0$ for all $\mu,\
x$. In particular, let $U$ be a $C^{\infty}$ vectorfield on $\cal
M$. Then each embedding $X$ determines an element, $x\mapsto
U^{\mu}(X(x))\,(\partial/\partial y^\mu)_{X(x)}$, of $T_X{\cal E}$.
This is the type of smearing used in \cite{I+K}.

	Let us calculate the gradient of ${\cal H}_N$. Starting from
Eq.\ (\ref{cst1}), using Eqs.\ (\ref{H1}), (\ref{H2}) and (\ref{H3}), and
integrating by parts, we obtain
\begin{eqnarray}
    {\text{grad}}_{\varphi}{\cal H}_N & = &
		(\det\gamma)^{1/2}\,[-(N^{\mu} L^k_{\mu})_{\| k} - m^2
		\varphi\,n_{\mu}N^{\mu}], 		\label{compFi} \\
	{\text{grad}}_{\pi}{\cal H}_N 	& = &
			L^{\bot}_{\mu}N^{\mu}, 		\label{compPi} \\
	({\text{grad}}_X {\cal H}_N )_\nu & = &
		N^{\mu}_{\|\nu}{\cal H}_{\mu}-\frac{1}{2}
		(\det\gamma)^{1/2}(N^{\mu}K^k_{\mu\nu})_{\| k} - P_{\kappa}
		\Gamma^{\kappa}_{\mu\nu}N^{\mu},	\label{compX}  \\
	({\text{grad}}_P {\cal H}_N)^{\nu} & = & N^{\nu}, \label{compP}
\end{eqnarray}
where $L^{\bot}_{\mu},L^k_{\mu}$ and $K^k_{\mu\nu}$ are given by
Eqs.\ (\ref{L1}), (\ref{L2}) and (\ref{K}). In \cite{I+K}, the
following theorem was shown:
\begin{thm}
     Let $M$ and $N$ be two $C^{\infty}$ vector fields
     on $\cal M$. Then
     \begin{equation}
          \{{\cal H}_M,{\cal H}_N\} = -{\cal H}_{[M,N]} , \label{poisscon}
     \end{equation}
     where $[M,N]$ is the Lie bracket of the fields $M^{\mu}$ and
     $N^{\mu}$.
\end{thm}

	Substituting the expressions Eqs.\ (\ref{compFi})--(\ref{compP}) for
the gradients into the Poisson brackets (\ref{poisscon}) we obtain
an identity that plays an important role in some proofs:
\begin{eqnarray}
     \lefteqn{\frac{1}{2}\int_{\Sigma}d^3 x\,(\det \gamma )^{1/2}
     [(M^{\mu}K^k_{\mu\nu})_{\| k}N^{\nu} -
     (N^{\mu}K^k_{\mu\nu})_{\| k}M^{\nu} ] =} \nonumber \\
     & & \quad\int_{\Sigma}d^3 x\,({\text{grad}}_{\varphi}
			{\cal H}_M\,{\text{grad}}_{\pi}{\cal H}_N
    - {\text{grad}}_{\pi}{\cal H}_M\,{\text{grad}}_{\varphi}{\cal H}_N).
								\label{I+Kid}
\end{eqnarray}

	The following theorem was essentially shown in \cite{I+K}:
\begin{thm}
     Let $\phi$ satisfy the Klein-Gordon equation (\ref{K-G}) on
	$\cal M$. Then
     for each curve $\lambda\mapsto X_{\lambda}$ with the tangent vector field
     $N$ on $\cal E$, the initial data $(\varphi_{\lambda},
     \pi_{\lambda})$ for $\phi$ on $X_\lambda(\Sigma)$ satisfy the evolution
	equation
     \begin{equation}
          (\dot{\varphi},\dot{\pi},\dot{X},\dot{P}) =
			J({\text{grad}}{\cal H}_N).           \label{evoleq}
     \end{equation}

     The equation for $\dot{P}$ is a consequence of the first three
     equations and the constraints ${\cal H}_N = 0$ for all
	$N \in T_X{\cal E}$.

	Conversely, if a curve
	$\lambda\mapsto(\varphi_{\lambda},\pi_{\lambda}, X_{\lambda})$
     on $\Gamma_{\phi}\times \cal E$ satisfies the evolution equations
     (\ref{evoleq}), then it defines a unique solution $\phi$ of the
     Klein-Gordon equation.
\end{thm}
Thus the Hamiltonian vector fields $J({\text{grad}}{\cal H}_N)$ of
the functions ${\cal H}_N$ are tangential to $\tilde{\Gamma}$, and
hence the system is first class, according to the definition given
in \cite{timelevels}.

	A simple consequence of theorem 4 is that the pull-back of the vector
field (\ref{evoleq}) to $\Gamma_{\phi}\times\cal E$ is given by
\begin{eqnarray}
     \dot{\varphi} & = & {\text{grad}}_{\pi}{\cal H}_N , \label{evoleqFi} \\
     \dot{\pi} & = & -{\text{grad}}_{\varphi}{\cal H}_N , \label{evoleqPi} \\
     \dot{X} & = & N. \label{evoleqX}
\end{eqnarray}

	Let us denote the space of longitudinal vectors at
$(\varphi,\pi,X) \in \Gamma_{\phi}\times \cal E$ by
$\Xi_{(\varphi,\pi,X)}$, {\em i.e.},
\begin{equation}
     \Xi_{(\varphi,\pi,X)} := \{(\Phi,\Pi,V)\in T_{(\varphi,\pi,X)}
		\tilde{\Gamma}\mid 	\Phi = {\text{grad}}_{\pi}{\cal H}_N,
				\Pi = -{\text{grad}}_{\varphi}{\cal H}_N,
					V = N\in T_X{\cal E}\}. \label{Xi}
\end{equation}
The map
$({\text{grad}}_{\pi}{\cal H}_N, -{\text{grad}}_{\varphi}{\cal H}_N):
T_X{\cal E}\rightarrow T_{(\varphi,\pi)}\Gamma_{\phi}$
is continuous and hence, by the closed graph theorem,
$\Xi_{(\varphi,\pi,X)}$ is a {\em closed\/} subspace of
$T_{(\varphi,\pi,X)}\tilde{\Gamma}$.

	Another consequence of theorem 4 is that the Fr\'{e}chet derivative of
the map $\rho_{XX'}$ with respect to $X'$ is given by
\begin{equation}
     {\text{D}}_{X'}\rho_{XX'}|_{(\varphi,\pi)}(V)=
     ({\text{grad}}_{\pi}{\cal H}_V(\varphi',\pi',X'),-{\text{grad}}_{\varphi}
     {\cal H}_V(\varphi',\pi',X')), \label{derrho}
\end{equation}
where $(\varphi',\pi') := \rho_{XX'}(\varphi,\pi)$.

	We shall need the pull-back of the symplectic form to the constraint
manifold. This is given by the following theorem.
\begin{thm}
     The pull-back $\tilde{\Omega}$ of the form $\Omega$ is given by the
     formula
     \begin{eqnarray}
          \lefteqn{\tilde{\Omega}((\Phi_1,\Pi_1,V_1 ),
          (\Phi_2,\Pi_2,V_2)) =} \nonumber \\
          &&\int_\Sigma d^3x\,[(\Pi_1+{\text{grad}}_{\varphi}{\cal H}_{V_1})
          (\Phi_2 - {\text{grad}}_{\pi}{\cal H}_{V_2}) -
          (\Pi_2 + {\text{grad}}_{\varphi}{\cal H}_{V_2})
          (\Phi_1 - {\text{grad}}_{\pi}{\cal H}_{V_1})]. \label{pulback}
     \end{eqnarray}
\end{thm}
{\bf Proof} The pull-back by the map $\tilde{C}$ of the form $\Omega$ is
given by
\begin{equation}
	\tilde\Omega((\Phi_1,\Pi_1,V_1),(\Phi_2,\Pi_2,V_2)) :=
     \Omega({\text{D}}\tilde{C}(\Phi_1,\Pi_1,V_1),
			{\text{D}}\tilde{C}(\Phi_2,\Pi_2,V_2)).
\end{equation}
Substituting into this equation the expressions for
${\text{D}}\tilde{C}$ from Eqs.\ (\ref{DC}) and (\ref{DHphi}), and
using Eq.\ (\ref{I+Kid}), one easily arrives at Eq.\
(\ref{pulback}). \hfill {\bf QED}

	Thus, $\tilde{\Omega}$ is degenerate, and the degeneracy subspace at
the point $(\varphi,\pi,X)\in\tilde{\Gamma}$ coincides with
$\Xi_{(\varphi,\pi,X)}$.

	The last important notion involving the constraint submanifold
is that of a `$c$-orbit', defined to be the set of points in $\tilde{\Gamma}$
that correspond to just one maximal classical solution (see
\cite{timelevels}). Let
$\tilde{\gamma}_{(\varphi,\pi,X)}$ be the map
$\tilde{\gamma}_{(\varphi,\pi,X)} : {\cal E} \rightarrow \Gamma_{\phi}
\times {\cal E}$ defined by
\begin{equation}
     \tilde{\gamma}_{(\varphi,\pi,X)}(X') := (\rho_{XX'}(\varphi ,\pi ),
     X')
\end{equation}
Then the $c$-orbit $\gamma_{(\varphi,\pi,X)}$ through the point
$(\varphi,\pi,X)\in\Gamma_{\phi} \times {\cal E}$ is defined as
\begin{equation}
     \gamma_{(\varphi,\pi,X)}:= \tilde{\gamma}_{(\varphi,\pi,X)}
     ({\cal E})
\end{equation}
{\em i.e.}, $\gamma_{(\varphi,\pi,X)}$ is the collection of all
embeddings, and Cauchy data on such, induced by the unique solution to
the field equations whose Cauchy data on $X(\Sigma)$ is $(\varphi,\pi)$.

    We shall show that the $c$-orbits are smooth submanifolds of
$\tilde{\Gamma}$ and that their tangent spaces coincide with
$\Xi_{(\varphi,\pi,X)}$; the proof is analogous to that of theorem
2.

	The tangent space to $\gamma$ at $\tilde{\gamma}_{(\varphi,\pi,X)}(X')$
is the image of the map
${\text{D}}\tilde{\gamma}_{(\varphi,\pi,X)}|_{X'} : T_{X'}{\cal E}
\rightarrow T_{(\varphi',\pi',X')}(\Gamma_{\phi}\times{\cal E})$,
where $(\varphi',\pi') :=\rho_{XX'}(\varphi,\pi)$.
Using Eq.\ (\ref{derrho}) we obtain, for all $V\in T_{X'}{\cal E}$,
\begin{eqnarray}
     {\text{D}}\tilde{\gamma}_{(\varphi,\pi,X)}|_{X'}(V) & = &
     ({\text{D}}\rho_{XX'}(\varphi,\pi )|_{X'}(V),V)  \nonumber \\
     & = & ({\text{grad}}_{\pi'}{\cal H}_V(\varphi',\pi',X'),
     -{\text{grad}}_{\varphi'}{\cal H}_V(\varphi',\pi',X'),V).
     \label{Dgamma}
\end{eqnarray}
Hence, ${\text{D}}\tilde{\gamma}_{(\varphi,\pi,X)|_{X'}}$ is injective,
and a comparison of Eq. (\ref{Dgamma}) with Eqs.\
(\ref{evoleqFi}--\ref{evoleqX}) shows that ${\text{D}}
\tilde{\gamma}_{(\varphi,\pi,X)}|_{X'}(T_{X'}{\cal E}) =
\Xi_{(\varphi,\pi,X)}$. As $\Xi$ is a closed subspace of a Hilbert
space, it splits, and the claims above are proved.

\section{Conclusions}
We have shown that, with proper functional-analytical care, the
geometrical structure of an infinite-dimensional parametrized system
can be developed in a way that is analogous to that of a
finite-dimensional system. In particular, for the model considered,
the extended phase space is a (weak-)symplectic infinite-dimensional
manifold, the constraint set is a submanifold of the phase space,
and the $c$-orbits are submanifolds of the constraint set. The
criteria for a constrained system to be first class are of the same
form as those of a finite-dimensional system. Many constructions
available for a finite-dimensional system can now be performed in
the infinite-dimensional case. The only difference is that the
symplectic form is only weakly non-degenerate. However, physicists
usually work with a restricted class of functions so that the
Poisson brackets are still well-defined.

	 We anticipate that our main results are broadly generalizable.
For example, an extension to the case where the Cauchy hypersurfaces
are asymptotically flat is likely to be relatively straightforward.
We hope that our results will be useful for a number of purposes. In
particular, our main goal was to apply the perennial formalism to
the scalar field system; this is be done in the accompanying paper.

\subsection*{Acknowledgements}
P.H. thanks the Theoretical Physics Group at Imperial College for
their hospitality. Both authors gratefully acknowledge
financial support from the European Network {\em Physical and
Mathematical Aspects of Fundamental Interactions}.

\appendix
\section{Cauchy problem}
We collect together some well-known results about the Cauchy
problem of linear hyperbolic systems and then use them to sketch a proof
of the theorem 1 in section \ref{Sec:dynamics}.

First, we state some lemmas about the space-time $({\cal M},g)$.
\begin{lem}
\label{coord}
     Let $X$ and $X'$ be two embeddings in $\cal E$ such that
     $X'(\Sigma) \subset I^+(X(\Sigma))$.
     Then, given $T>0$ and $\epsilon>0$,
     there is a one-dimensional family $\{X_t\}$, $t \in (-\epsilon,
     T+\epsilon)$, of embeddings such that:
     \begin{description}
          \item[a)] $X_0 = X,\quad X_T(\sigma) = X'(\Sigma)$;

          \item[b)] if $(U,h)$ is a chart of $\Sigma$ ($h(U) \subset
               \mathR^3$), then $(\bigcup_{t\in(-\epsilon,T+\epsilon)}X_t (U),
               (X_t(h^{-1}(x)))^{-1})$
               is a $C^{r+1}$ chart in $\cal M$, where $X_t(h^{-1}(x))$ is
               considered as a map
               \begin{equation}
                    X_t(h^{-1}(x)) : (-\epsilon,T+\epsilon) \times h(U)
                    \rightarrow {\cal M};
               \end{equation}
          \item[c)] the components $g_{\alpha\beta}(t,x)$ of the metric in
               any such chart satisfy the equations
               \begin{equation}
                    g_{00}(t,x)<0
		\label{A12.1}
               \end{equation}
               in $(-\epsilon,T+\epsilon) \times h(U)$ (the $t$-curves are
               everywhere timelike), and
               \begin{equation}
               \begin{array}{lclcl}
                    g_{00}(0,x) & = & g_{00}(T,x) & = & -1, \\
                    g_{0i}(0,x) & = & g_{0i}(T,x) & = & 0,
               \end{array}
               \end{equation}
               for all $x\in h(U)$.
     \end{description}
\end{lem}
The proof is simple. Observe that $g^{00}$ is negative and $g_{kl}$ is
positive-definite everywhere in $(-\epsilon,T+\epsilon) \times
h(U)$; this is because any $X_t(U)$ hypersurface is a part of a Cauchy
hypersurface, and hence spacelike. Then the condition that $g^{kl}$ is also
positive-definite everywhere is equivalent to Eq.\ (\ref{A12.1}). Hence, for
any $t\in[0,T]$ and $x\in h(U)$, we have
\begin{eqnarray}
     |g^{00}| & > & c_1,
	\label{A12.2} \\
     g^{kl}\xi_k \xi_l & > & c_2 e^{kl}\xi_k\xi_l, \label{A12.3}
\end{eqnarray}
where $c_1$ and $c_2$ are positive constants, $e_{kl}$ is a
positive-definite $C^{\infty}$ metric on $\Sigma$, and $\xi_k$ is an arbitrary
covector field on $\Sigma$.

The proof of the following lemma can be found in \cite{H+E}:
\begin{lem}
\label{futureS}
     If $\Sigma_1$ and $\Sigma_2$ are two Cauchy hypersurfaces in $(M,g)$
     then there is a Cauchy hypersurface $\Sigma_3$ such that
     \begin{equation}
          \Sigma_3 \subset (I^+ (\Sigma_1) \cap
          I^- (\Sigma_2)). \label{A13.1}
     \end{equation}
\end{lem}
Thus, $\Sigma_1 \cap \Sigma_3 =\Sigma_2 \cap \Sigma_3 = \emptyset$.

	Next, consider the Klein-Gordon equation (\ref{K-G}) for the
field $\phi$, and the associated Cauchy problem. In the chart
described in lemma \ref{coord}, Eq.\ (\ref{K-G}) has the form
\begin{eqnarray}
     -g^{00}\frac{\partial^2\phi}{\partial t^2} & = &
     g^{kl}\frac{\partial^2\phi}{\partial x^k\partial x^l} +
     2g^{0k}\frac{\partial^2\phi}{\partial t\partial x^k} +
     |\det g|^{-1/2}\,\partial_{\mu}(|\det g|^{1/2}g^{0\mu})\frac{\partial
     \phi}{\partial t} \nonumber \\ & + &
     |\det g|^{-1/2}\,\partial_{\mu}(|\det g|^{1/2}g^{k\mu})\frac{\partial
     \phi}{\partial x^k} + m^2 \phi.  \label{A13.3}
\end{eqnarray}
An initial datum for $\phi$ at the time $t$ is the pair of scalar
fields $(\varphi_t,\dot{\varphi}_t)$ on $\Sigma$ given by
\begin{eqnarray}
     \varphi_t(x) & = & \phi(t,x), \\
     \dot{\varphi}_t(x) & = & \frac{\partial \phi}{\partial t},
\end{eqnarray}
where $\varphi_t(x)$ coincides with the Cauchy datum for $X_t(\Sigma)$ as
defined by Eq.\  (\ref{Def:pi}). For $t=0$ and $t=T$, we have also
\begin{eqnarray}
     (\det \gamma)^{1/2}\,\dot{\varphi}_0 (x) & = & \pi_0 (x), \\
     (\det \gamma)^{1/2}\,\dot{\varphi}_T (x) & = & \pi_T (x),
\end{eqnarray}
where $\pi_0 (x)$ and $\pi_T (x)$ are the pieces of Cauchy data defined
by Eq.\ (\ref{Def:varphi}) for $X_0(\Sigma)$ and $X_T(\Sigma)$.

Consider the coefficient functions of Eq.\ (\ref{A13.3}). They define
tensor fields on $\Sigma$. Indeed,
\begin{eqnarray}
     a^{00}(t,x) & = & -g^{00}(t,x), \\
    a^0(t,x) & = &  |\det g|^{-1/2}\,\partial_{\mu}(|\det g|^{1/2}g^{0\mu}
     )|_{t,x},\\
     a(t,x) & = & m^2,
\end{eqnarray}
are scalar fields on $\Sigma$ for each $t$, whereas
\begin{eqnarray}
     a^{0k}(t,x) & = & g^{0k}(t,x), \\
     a^k(t,x) & = & |\det g|^{-1/2}\,\partial_{\mu}(|\det g|^{1/2}g^{k\mu}
     )|_{t,x},
\end{eqnarray}
are contravariant vector fields for each $t$, and
\begin{equation}
     a^{kl}(t,x) = g^{kl}(t,x)
\end{equation}
is a contravariant tensor of second rank on $\Sigma$ for each $t$.

	The $C^{r+1}$-differentiability of $X$ implies the following
properties of these tensor fields:
\begin{equation}
\begin{array}{lclcl}
     a^{\alpha\beta} & \in & {\text{Lip}}([0,T];H_0^{r-1}(\Sigma)) & \subset
     & L^{\infty}([0,T];H_0^r(\Sigma)), \\
     a^{\alpha} & \in & {\text{Lip}}([0,T];H_0^{r-2}(\Sigma)) & \subset
     & L^{\infty}([0,T];H_0^{r-1}(\Sigma)), \\
     a & \in & {\text{Lip}}([0,T];H_0^{\infty}(\Sigma)) & =
     & L^{\infty}([0,T];H_0^r(\Sigma)).
\end{array}
\end{equation}
Moreover, according to Eqs.\ (\ref{A12.2}) and (\ref{A12.3}),
we have $a^{00}(t,x) \geq c_1$, for all $t\in [0,T]$ and $x \in \Sigma$, and
\begin{equation}
     a^{kl}(t,x)\xi_k (x)\xi_l (x) \geq c'_2 e^{kl}(x)\xi_k (x)\xi_l (x),
\end{equation}
for all $t \in [0,T]$ and $x \in \Sigma$. The relations above enable us
to use `localized' forms of the theorems 4.15 and 4.13 in \cite{F+M}, and
to apply them to construct unique local evolution systems $F_{t,s}$ which
can be `patched together'. By this means we are able to prove the
following lemma (for further details see \cite{F+M}):
\begin{lem}
\label{evol.syst}
     Let the assumptions of Lemma \ref{coord} be satisfied, and let
$r\geq 4$.  Then, each initial datum $(\varphi_0 (x),\dot{\varphi}_0
(x))\in H_0^{s+1}(\Sigma)\times H_0^s(\Sigma)$ on $X_0 (\Sigma)$, where
$1 \leq s \leq r-1$, defines a unique solution $\phi$ to the equation
(\ref{K-G}) in $U$ whose initial datum $(\varphi_T
(X),\dot{\varphi}_T(x))$ on $X_T (\Sigma)$ belongs to the Sobolev space
$(\varphi_T (X),\dot{\varphi}_T(x)) \in H_0^{s+1}(\Sigma)\times H_0^s
(\Sigma)$. Furthermore, the maps $U(0,T):H_0^{s+1}(\Sigma)\times H_0^s(\Sigma)
\rightarrow H_0^{s+1}(\Sigma)\times H_0^s(\Sigma)$ are
automorphisms of Banach spaces.
\end{lem}
Note that it is a trivial matter to pass from the initial data
$(\varphi,\dot{\varphi})$ to the Cauchy data $(\varphi,\pi)$: since
$(\det\gamma)^{1/2}$ is $C^{\infty}$ and bounded below by zero in $U$,
the definition $\pi (x) := (\det \gamma)^{1/2}\,\dot{\varphi}(x)$
describes an isomorphism between the Banach spaces $H_0^s(\Sigma)$ and
$H_1^s(\Sigma)$ for any $s \leq r$.

	Finally, given any pair of arbitrary embeddings $X_1$ and $X_2$, we
can use lemma \ref{futureS} to find an `intermediate' embedding $X_3$. To
be able to apply lemma \ref{coord} to the pairs $\{X_1,X_3\}$ and
$\{X_3,X_2\}$, we must find two embeddings $X'_3$ and $X''_3$ such that
the corresponding $t$-curves are timelike, and with $X'_3 (\Sigma) =
X''_3 (\Sigma) = X_3 (\Sigma)$. Then, we can use lemma \ref{evol.syst}
and the diffeomorphism invariance of $B$ to prove the theorem 1. \hfill
{\bf QED}



\begin{thebibliography}{99}

\bibitem{dewitt} B.~S.~DeWitt, {\em Phys.\ Reports} {\bf 19C}
295--357 (1975).

\bibitem{isham-texas} C.~J.~Isham, ``Quantum field theory in curved
	space-time, an overview.'' In {\em Proceedings of the Eigth
	Texas Symposium on Relativistic Astrophysics}, ed.  M.~Papagiannis,
	New York Academy of Sciences, New York, pp 114--157 (1977).

\bibitem{Hawking} S.~W.~Hawking, {\em Comm.\ Math.\ Phys.} {\bf 43}
	199--220 (1975).

\bibitem{K45} P.~A.~M.~Dirac, {\em Can.\ J.\ Math.} {\bf 3} 1--14
	(1951).

\bibitem{hyperspace} K.~V.~Kucha\v{r}, {\em J.\ Math.\ Phys.}
	{\bf 17} 777--791 (1976).

\bibitem{K47} K.~V.~Kucha\v{r}, {\em J.\ Math.\ Phys.} {\bf 17}
	801--820 (1976).

\bibitem{I+K} C.~J.~Isham and K.~V.~Kucha\v{r}, {\em Ann.\ Phys.} {\bf 164}
    288--315 (1985).

\bibitem{K33} K.~V.~Kucha\v{r}, ``Canonical methods of
	quantization'', in {\em Quantum Gravity 2: A Second Oxford
     Symposium}, eds. C.~J.~Isham, R.~Penrose and D.~W.~Sciama,
	Clarendon University Press, Oxford, pp 329--374 (1981).

\bibitem{dirac} P.~A.~M.~Dirac, {\em Rev.\ Mod.\ Phys.} {\bf 21}
	392--399 (1949).

\bibitem{timelevels} P.~H\'{a}j\'{\i}\v{c}ek,
	{\em J.\ Math.\ Phys.} {\bf 36} 4612--4638 (1995).

\bibitem{isham}
     C.~J.~Isham, ``Topological and global aspects of quantum
	theory'' in {\em Relativity, Groups and Topology II},
    eds. R.~Stora and B.~S.~DeWitt, North-Holland, Amsterdam,
	pp 1061--1290 (1984).

\bibitem{blue-book} A.~Ashtekar, {\em Lectures on Non-Perturbative Quantum
     Gravity}, World Scientific, Singapore (1991).

\bibitem{C+M} P.~R.~Chernov and J.~E.~Marsden, {\em Properties of
     Infinite Dimensional Hamiltonian Systems}, Lecture Notes in
     Mathematics, eds. A.~Dold and B.~Eckmann. Springer-Verlag, Berlin
	(1974).

\bibitem{H+I} P.~H\'{a}j\'{\i}\v{c}ek and C.~J.~Isham, ``Perennials
	and the group-theoretical quantization of a parametrized scalar
	field on a curved background'', Imperial College preprint
	IMPERIAL/TP/95--96/2 (1995).

\bibitem{F+M} A.~E.~Fischer and J.~E.~Marsden, ``The initial value
	problem and the dynamical formulation of general relativity'' in
	{\em General Relativity.  An Einstein Centenary Review}, eds.
	S.~W.~Hawking and W.~Israel, Cambridge University Press,
	Cambridge, pp 138--211 (1979).

\bibitem{H+E} S.~W.~Hawking and J.~F.~R.~Ellis, {\em The Large Scale
     Structure of Space-Time}, Cambridge University Press, Cambridge
     (1973).

\bibitem{G+V} I.~M.~Gel'fand and N.~J.~Vilenkin, {\em Generalized
     Functions IV}, Academic Press, New York (1964).

\bibitem{kuch-prehled} K.~V.~Kucha\v{r}, ``Time and interpretations of
    quantum gravity'', in {\em Proceedings of the 4th Canadian
	Conference on General Relativity and Relativistic Astrophysics},
	World Scientific, Singapore, pp 211--314 (1992).

\bibitem{eells} J.~Eells, ``On the Geometry of Function Spaces'', in {\em
	Symposium de Topoloqia Algebrica},
	303--307 Mexico (1958).

\bibitem{M+E+F} J.~E.~Marsden, D.~G.~Ebin and A.~E.~Fischer
	``Diffeomorphism Groups, Hydrodynamics and Relativity'',
	in {\em Proceedings of the Thirteenth Biennial Seminar of the Canadian
     Mathematical Congress}, ed. J.~R.~Vanstone, Canadian Mathematical
     Congress, Montreal, pp 135--279 (1972).

\bibitem{Lang} S.~Lang, {\em Differential Manifolds}, Addison-Wesley,
     Reading (1972).

\end{thebibliography}
\end{document}